\documentclass{article} 
 
\begin{document} 
\begin{center} 
\title{Quantum mechanical observer and superstring/M theory} 
\author{M.Dance}
\maketitle
\end{center} 
 
\begin{abstract} 
Terms are suggested for inclusion in a Lagrangian density as seen by an observer O2, to represent the dynamics of a quantum mechanical observer O1 that is an initial stage in an observation process.  This 
paper extends an earlier paper which suggested that the centre-of-mass kinetic energy of O1 could correspond to, and possibly underlie, the Lagrangian density for bosonic string theory, where the worldsheet 
coordinates are the coordinates which O1 can observe.  The present paper considers a fermion internal to O1, in addition to O1's centre of mass.  It is suggested that quantum mechanical uncertainties 
in the transformation between O1's and O2's reference systems might require O2 to use $d$ spinor fields for this fermion, where $d$ is the number of spacetime dimensions.  If this is the case, and if the 
symmetry/observability arguments in~\cite{Dance0601} apply, the resulting Lagrangian density for the dynamics of O1 might resemble, or even underlie, superstring/M theory.  
\end{abstract} 
 
\section{Introduction}
 
String theory dualities appear to point to the existence of an as yet unknown unifying theory, dubbed M-theory.  The underlying physical basis of M theory is unknown, and may depend on concepts as yet 
undiscovered~\cite{Schwarz0702}.  
 
A recent paper~\cite{Dance0601} suggested that a simple Lagrangian density for a quantum mechanical observer O1 - its centre-of-mass kinetic energy as understood by another observer O2 - might in some 
circumstances resemble the Lagrangian density for bosonic string theory.  Such an observer O1 could perhaps arise as an initial observation stage, within an overall observation process in which O2 is a 
subsequent observation stage.  In~\cite{Dance0601}, the worldsheet coordinates of the string theory are simply those spacetime coordinates which O1 can observe in its own reference system.  It was there 
suggested that the dynamics of a quantum mechanical observer such as O1 (and the nature of quantum mechanical transformations between reference systems) might fundamentally underlie string theory.  A good 
introductory review of the developing field of quantum mechanical reference systems and their transformations can be found in~\cite{Bartlett0610}.
 
The present paper briefly reviews~\cite{Dance0601} and extends it by including simple internal dynamics for the observer O1.  A possible connection with superstring/M theory is suggested.
 
To some extent the material in the present paper, and in~\cite{Dance0601}, might call to mind the AdS/CFT correspondence for cold atoms~\cite{Son0804}, \cite{BM0804}.  The AdS/CFT correspondence is not 
applied or discussed below; the present paper proceeds along a different line, but it seems possible that it could be connected in some way to the correspondence.   
 
Other authors have discussed atomic Zitterbewegung (trembling centre of mass motion) in relation to relativistic dynamics of cold atoms,e.g.~\cite{Merkl0803}.  Some authors may have made a recent suggestion that Zitterbewegung could be related to string theory, similar to~\cite{Dance0601}.  The existence of Zitterbewegung has been challenged elsewhere as what one could perhaps consider an artefact, or at least a subsidiary phenomenon arising from non-commutative geometry.  In fact, all of these effects might be connected to quantum mechanical observers, and each perspective might be equally as valid as every other, 
representing different descriptions of the same physics.
 
Other work, in the general area of relating atoms to gravity and high energy physics, includes the string-net concept, in which an extended condensed matter (spin) system might have the potential to explain the origin of elementary particles~\cite{Wen03}, and can be related to loop quantum gravity~\cite{Gu&Wen06}.

\section{Review - O1 CM and bosonic string theory}
 
In~\cite{Dance0601}, it was suggested that the centre-of-mass (CM) kinetic energy density of a quantum mechanical observer O1 could be included in the Lagrangian density $\mathcal{L}$ which another observer 
(later in the overall observation process) "sees".  In some circumstances, described in~\cite{Dance0601}, such an $\mathcal{L}$ might correspond to the Lagrangian density for bosonic string theory. 
 
Why add observer terms to Lagrangian densities?  Many observations occur in a number of stages.  There is an initial event, often quantum mechanical, followed by amplification and processing.  The initial 
quantum mechanical event may involve a single atom or molecule as a quantum mechanical observer.  For many years, workers have modelled quantum measurement by Hamiltonians with terms for observed systems and 
measuring apparatus.  It seems fair to include observers in Lagrangians, as appropriate.  
 
\cite{Dance0601} considered the Lagrangian (density) of a quantum mechanical observer O1 as seen (or known) by an observer/observation stage O2.  The Lagrangian (density) was simply the nonrelativistic kinetic energy of O1's centre of mass:-
\begin{equation}
L^{EK}_{obs} = \frac{1}{2}m \eta_{ij}\frac{dX^i}{dT}\frac{dX^j}{dT}
\end{equation} 
where $m$ is the mass of O1, and \{$X^{\mu }$\} are fields which represent the position of O1's centre of mass in O2's coordinates (and $T = X_{0}$).  The independent coordinates were then changed to those of O1's reference 
system:-  
\begin{equation}
L^{EK}_{obs} = \frac{1}{2}m \kappa^{\alpha \beta } \eta_{\mu \nu } 
                        \frac{\partial{X^\mu}}{\partial{x^\alpha}}  
                        \frac{\partial{X^\nu}}{\partial{x^\beta}}
\end{equation} 
where $x^{\alpha }$ and $x^{\beta }$ are the coordinates ($t$, $x$, $y$, $z$,..) internal to the observer O1.  $\kappa $ is a coordinate-dependent factor which depends on the transformation between the $x^
{\alpha } $ and $X^{\mu } $.   $\kappa$ can be divided by an appropriate volume factor so that the expression above becomes a Lagrangian density $\mathcal{L}^{EK}_{obs}$ .

It might seem strange to think of an observer's centre of mass kinetic energy in that observer's own internal coordinates.  Quantum mechanically, this need not be so surprising in view of the usual zero 
point energy.  Such thinking may also reflect the fact that O2's knowledge (or lack of it) about what O1 observes must incorporate quantum mechanical uncertainties in the O1-O2 reference system 
transformation.
 
\cite{Dance0601} then postulated that a simple quantum system O1 such as a single atom cannot generally measure its radial coordinate $r$ - transition amplitudes do not generally identify $r$.  \cite
{Dance0601} further postulated that symmetry of O1 may restrict O1's ability to observe the spherical polar coordinate $\theta$ and/or $\phi$.  The discussion in~\cite{Dance0601} concerned O1's 
observation of external fields, but the same kind of argument could equally apply to O1's own centre-of-mass kinetic energy.   (While these suggestions might seem unorthdox, it may be worth noting that Bohr, 
Ulfbeck and Mottelson~\cite{Bohr2004a}, \cite{Bohr2004b} have related the density matrix for a particular experimental outcome in nonrelativistic quantum mechanics to the spacetime symmetry group of the 
experimental configuration.) 
 
If O1 cannot measure $\theta$ or $r$, but can measure $t$ and $\phi$, then setting $\tau = t$ and $\sigma = \phi$, the Lagrangian density above turns into that of the simple bosonic string theory with 
worldsheet coordinates ($\tau$, $\sigma$), with a string tension related to the observer mass and worldsheet metric related to aspects of the O1-O2 coordinate transformation.
 
This has summarised \cite{Dance0601}.  We will now extend the argument of~\cite{Dance0601} to include an internal O1 fermion moving relative to O1's centre of mass.

\section{Adding internal O1 terms}
 
O1's centre of mass (CM) kinetic energy density is not where the story ends.  Other terms can be added, to represent O1's internal dynamics.  Typically O1 will be an atom, or interacting atoms in a molecule. 
 Let us assume that O1 is an nonrelativistic atom relative to O2, with an electron of interest for measurement purposes.  Then:
 
\begin{equation}
\mathcal{L}       =          \mathcal{L}_{\rm O1's CM, relative to O2}  
                                    +   \mathcal{L}_{\rm internal, relative to O1's CM}
\end{equation}
   
$\mathcal{L}_{\rm internal}$ could include the kinetic energy density of the electron relative to O1's centre of mass, a potential energy term, and an electron rest mass term.   Which of these will be of 
most interest to us?  
 
Let us choose (for now) the simplest possible term for the dynamics of O1's internal electron.  It is the Lagrangian density of a free, noninteracting massless fermion field, expressed as simply as possible in O2's coordinates:
\begin{equation}
\mathcal{L} _{\rm int} =  i\bar\psi \gamma ^{\nu} \partial_{\nu} \psi
\end{equation}
That is, in our simple model $\mathcal{L}$ (as seen by O2), is O1's centre of mass kinetic energy density, plus the kinetic energy density of the fermion motion relative to (the quantum expectation value of 
the position of) O1's centre of mass. 

The independent coordinates are those of O2's reference system.  We will now transform them to O1's reference system.  This was done for the first term in~\cite{Dance0601}.  It will be done for the second 
term below.
 
The present paper assumes that O1 is nonrelativistic relative to O2.  A better formulation would be relativistically covariant.  It is notable that the formulation of a relativistically covariant theory of 
quantum mechanical bound states is an active area of research, recently described by two authors as an open problem ~\cite{Kim&Noz0609} - see also~\cite{Kim&Noz08} for recent further work in this area.  So 
choices for $\mathcal{L}$ may to some extent be subject to further developments in this general area.

\section{Transforming independent coordinates to O1 system}
 
In~\cite{Dance0601}, the first term in $\mathcal{L}$ above was transformed to:
\begin{equation}
\mathcal{L}_{\rm CM} =          \frac{1}{2}m \kappa^{\alpha \beta } \eta_{\mu \nu } 
                                    \frac{\partial{X^\mu}}{\partial{x^\alpha}}  
                                    \frac{\partial{X^\nu}}{\partial{x^\beta}} 
\end{equation}
where the $x^{\alpha}$, $x^{\beta}$ are coordinates in O1's reference system.  
 
Let us now consider the second term in $\mathcal{L}$.  If we were dealing with a nice, clean transformation between classical reference frames, we would simply have:
\begin{equation}
\mathcal{L}_{\rm internal}         =          i\bar\psi \gamma ^{\alpha} \partial_{\alpha} \psi
\end{equation}
because  $\gamma ^{\nu} \partial_{\nu} =  \gamma ^{\alpha} \partial_{\alpha}$.
 
However, the situation here is not so straightforward, because we have to consider a quantum mechanical transformation between the O2 and O1 reference systems, instead of a classical transformation.
 
Let us first consider quantum reference systems and transformations between them generally, before returning to handle the second term in $\mathcal{L}$.  The reader is also referred to e.g.~\cite
{Bartlett0610}, \cite{Wang0609}, and \cite{Dance0610} for further discussions about transformations between quantum mechanical reference systems.

\subsection{General discussion: transformation between quantum reference systems}
 
In an early paper on quantum mechanics~\cite{heisenberg}, Heisenberg noted that the uncertainty principle applies to observers as well as to observed systems.  That is, the quantum uncertainties in an 
observer O1's position and momentum, in another observer O2's coordinates, are subject to the uncertainty principle.  He noted that this could be a problem for relativity.  This is because the centrepiece of 
classical relativity - the Poincare transformation between two inertial reference frames - requires exact knowledge of an observer's position and velocity.  (Heisenberg postulated that any practical effect 
of observer indeterminacy could be eliminated by allowing the observer's mass to approach infinity.)  
 
There have been a number of approaches to making special relativity and quantum mechanics consistent, e.g. see~\cite{Kim&Noz0609}.  Some recent papers, e.g.~\cite{Wang0609} and \cite{Dance0610} have discussed 
Poincare transformations between reference frames when Heisenberg's point above is considered.  The central conclusion appears to be that one observer (e.g. O2) cannot know exactly how another observer O1 
describes spacetime.  O2 cannot be certain how to draw O1's coordinate axes - if O1 can still be said to have coordinate axes.  From the standpoint of what a given observer O2 can know, the nice coordinate 
axes drawn for O1 in classical spacetime diagrams become smeared out when quantum mechanical effects are included.  The effect becomes more marked when O1 has a high speed relative to O2.  Such 
transformations between observers' quantum mechanical reference systems may be best described as a quantum superposition of the usual (classical) Poincare transformations.  
 
The quantum mechanical transformation as a whole could be thought of as a cloud of vectors in ($p$,$x$) space (phase space).  In spacetime, for any given point in O2's reference system, one might also imagine 
a cloud of vectors - this time radiating from a central point in O1's reference system, as far as O2 knows, where that central point is O2's point that has undergone a simple classical Poincare 
transformation to a notional classical O1 reference frame.  (These uncertainties in a transformation can be thought of as those known by any one observer.  But the uncertainties might also apply to an 
observer's knowledge of even their own reference system.)
 
It must be emphasized here that the present paper concerns what O2 knows, and when expressed in O1's coordinates as the independent coordinates, we perhaps find out what O2 knows about what O1 sees.  By 
contrast, many papers on quantum mechanical reference system transformations appear to transform between what O1 knows and what O2 knows.  It seems that quantum mechanical transformations between reference 
systems might therefore fall into two distinct types: (1) (unitary) transformations between different observers, and (2) (nonunitary) transformations in the degree (or level) of inference made by a single 
observer O2 in relation to the observations of other observers.   The present paper may concern the latter type.  Arguably, this type is what we mainly do, as observers O2 ourselves.

\subsection{Transforming psi} 
Let us now consider what happens to the $\psi (x)$ field under a quantum mechanical transformation between the O2 and O1 reference systems.  
 
Standard quantum field theory describes field transformations under classical Poincare transformations. The field components as seen by a first observer become transformed, as seen by a second observer, into 
linear combinations of the first observer's components.  The coefficients reflect the type of field - scalar, spinor, vector, and so on.  
 
What happens to the spinor field $\psi$ under a quantum mechanical reference system transformation, instead of a classical Poincare transformation?  Can the quantum uncertainties in the transformation be 
absorbed into the spinor coefficients that arise in a classical transformation, or is something more required? 
 
As noted above, the quantum transformation may be expressible as a quantum superposition of classical Poincare transformations, e.g.~\cite{Wang0609}.  Overall, $\psi$ at $x$ in O2's reference system gets mapped (as far as O2 can know) to Poincare-transformed $\psi$ values in a spacetime region (or "cloud") in O1's coordinates (as far as O2 can know them).  
That is, the quantum mechanical transformation, as far as O2 can know it, maps O2's $\psi(x)$ at an individual point $x$ to a spinor field defined over a cloud of points in O1's coordinates, each point $q$ weighted with a quantum mechanical amplitude that describes how likely it seems to O2 that the point $x$ in O2 maps to $q$ in O1.  
However, standard quantum field theory treats observers using classical Poincare transformations, as if all observers know each others' reference frames with certainty.  
Let us now hypothesise that such a perspective will lump all of the uncertainties involved in quantum mechanical transformations between reference systems, instead into the fields. 
Then the "cloud of vectors" inherent in a quantum mechanical Poincare transformation would in some sense effectively be lumped into $\psi$.

Let us then postulate that to describe a fermion field in $d$ spacetime dimensions relativistically, $\psi(x)$ becomes a vector of spinors, having $d$ components $\psi^\mu (x)$ each of which is a spinor. In that case, $d$ spinor fields are needed to describe the dynamics of O1's internal fermion.  Then for relativistic covariance, both O1 and O2 would have to describe the fermion in this way.     

Another way of describing this postulate is to see that a quantum transformation, represented as a quantum superposition of classical transformations, would give rise to a quantum superposition in the mixing of the original spinor components of $\psi (x)$.  
It would be as if the mixing itself had quantum uncertainty. Instead of leaving this uncertainty with the mixing coefficients, we lump it into the fields, postulating that what we have here is equivalent to a theory in which spacetime transformations are classical, with postulated alterations in the behaviour and form of the relevant fields.  This aims to make a connection with standard quantum field theory, in which reference system transformations are classical.

\subsection{Transformed Lagrangian density}

The resulting form of the Lagrangian (density) for O1's dynamics is then: 
\begin{equation}
\mathcal{L}  =  \frac{1}{2}m \kappa^{\alpha \beta } \eta_{\mu \nu } 
                \frac{\partial{X^\mu}}{\partial{x^\alpha}}  
                \frac{\partial{X^\nu}}{\partial{x^\beta}} 
                + i\bar{\psi}^\mu \gamma ^{\alpha} \partial_{\alpha} \psi_\mu
\end{equation}

where the $x^{\alpha}$, $x^{\beta}$ are coordinates in O1's reference system.

\section{Removal of some O1 coordinates}
 
As in \cite{Dance0601}, let us postulate that a simple quantum mechanical O1 cannot generally measure (or pass information to O2 about) its radial coordinate $r$, as quantum mechanical interactions do not tend to pick out any particular $r$.  Then $r$ will drop out of the dynamics in an appropriate way, as if integrated out by O1.  As in \cite{Dance0601}, let us also postulate that if O1 has internal symmetry, this restricts O1's ability to detect/observe the relevant coordinate(s) ($\theta$ and/or $\phi$).

The resulting form of the Lagrangian (density) for O1's dynamics is then: 
\begin{equation}
\mathcal{L}       =          \frac{1}{2}m \kappa^{\alpha \beta } \eta_{\mu \nu } 
                        \frac{\partial{X^\mu}}{\partial{x^\alpha}}  
                        \frac{\partial{X^\nu}}{\partial{x^\beta}} 
                        + i\bar\psi^\mu \gamma ^{\alpha} \partial_{\alpha} \psi_\mu
\end{equation}
where the $x^{\alpha}$, $x^{\beta}$ are those coordinates in O1's reference system which O1 is capable of observing and relaying information about to O2.  I postulate that this will typically be a set of coordinates such as ($\tau$, $\sigma$) where $\tau$ is a time coordinate and $\sigma$ is an angular coordinate.  The {$\gamma^\alpha$} are the appropriate dimensionally-reduced version of the 4D originals.

$\mathcal{L} $ now has a form similar to the Lagrangian density of superstring/M theory.  For comparison, the reader is referred to the world-sheet superstring action given in~\cite{hep-ex/0008017} at page 25, in the conformal gauge:
\begin{equation}
S = -\frac{T}{2}\int d^2 \sigma (\partial_{\alpha} x^\mu \partial^{\alpha} x_\mu   -  i\bar{\psi}^\mu \rho^\alpha \partial_\alpha \psi_\mu )  
\end{equation}

It is tentatively suggested that the dynamics of the observer O1 and of the O2-O1 reference system transformation, as outlined above, might form (part of) a physical description underlying superstring/M theory, if the above line of argument is correct.

\section{Further comments}
 
\subsection{Key postulate remains to be shown}
A key postulate of the argument above was that for O2 to describe the fermion, using a Lagrangian density corresponding to a classical spacetime, $\psi$ would become a vector field of spinors, having $d$ components $\psi^{\mu}$ each of which is a spinor.  That is, it is postulated that $d$ fermion fields are now needed to describe the fermion dynamics.  It remains to be proven that this is the case.  
(However, if a quantum transformation of the spinor $\psi$ gives an integral over a "cloud" in spacetime, of an integrand involving a quantum amplitude weighting $\psi$ in the cloud, perhaps a Taylor expansion of $\psi$ about the cloud centre will give the $d$ spinors sought.)
 
\subsection{Radial coordinate}
The recent paper by Banerjee and Sen~\cite{hepth0805234} discusses how to interpret the M2-brane action.  These authors note that  the action contains a free field, decoupled from the other fields, which they interpret as the radial coordinate of the centre of mass of the brane.  By comparison, the $r$ which drops out in the present paper and in~\cite{Dance0601} is the radial coordinate of the observer O1's centre of mass, relative to the quantum expectation value of O1's centre of mass position.  In both cases, we have the radial coordinate of the centre of mass of the relevant entities.  The question might be (as above) whether M theory comes from a quantum mechanical observer.

\subsection{The fermion term}
In string theory, and in the line of argument of the present paper, it seems preferable (or even required) to include a fermion term to accompany the boson term.  (Observation with a detector generally requires a fermion field such as an electron in an atom, while the bosonic term for the observer's centre of mass motion may in some sense be inescapable.)  
  
\subsection{Feynman diagrams in superstring theory}
 
An introductory review~\cite{hep-ex/0008017} noted that:
 
\small
"In four of the five [consistent superstring theories] (heterotic and type II) the fundamental strings are oriented and unbreakable.  As a result, these theories have particularly simple perturbation expansions. Specifically, there is a unique Feynman diagram at each order of the loop expansion. .... For these four theories the unique L-loop diagram is a closed orientable genus-L Riemann surface, which 
can be visualised as a sphere with L handles.  External (incoming or outgoing) particles are represented by N points (or "punctures") on the Riemann surface."

\normalsize
The physical picture suggested in the present paper might explain the form of the Feynman diagrams (for four of the five consistent superstring theories) as follows.  The central sphere in the diagrams may represent the dynamics of the observer O1, including the 
uncertainty involved in the O1-O2 reference system transformation.  The handles in the diagrams might represent O1's interaction with itself or with atomic observers indistinguishable from O1 (as far as O2 knows).
 
Why must the lines for external particles enter and leave the surface at puncture points?  External particles are from an observed system S.  In the picture in the present paper, $\mathcal{L}$ contains no terms (as yet) for S; S still awaits description; but what we \emph{can} say for now is that everything that the observer O2 knows about S is filtered through O1.   The requirement that all external lines begin or end on the central surface could correspond to the fact that all of O2's measurements come via O1.  This should be a good approximation provided that the S-O1 interaction and the O1-O2 reference system transformation produce much stronger quantum effects than any direct S-O2 interaction does.

\subsection{Alternative physical picture -- an observed system S}
 
It is interesting to consider whether a better physical explanation might instead be that the $X^\mu $ fields represent coordinates of an observed system S, possibly the centre of mass position of S, rather than the centre of mass position of O1.  The independent coordinates in the final expression for the Lagrangian density would still be those describing the reference system of the quantum mechanical observer O1.  However, it might be more difficult to apply a symmetry principle to eliminate some of the O1 coordinates in this picture.  It might also turn out to be more difficult to explain dualities (e.g. S-duality) in this alternative picture.

\section{Conclusion}    
 
In the present draft paper, some simple terms for a quantum mechanical observer O1 have been included in a Lagrangian density as seen by a later-stage observer O2.  An earlier paper~\cite{Dance0601} suggested that a kinetic energy density term for a first-stage quantum observer's centre of mass might correspond to the Lagrangian density of bosonic string theory.  The present paper extends~\cite{Dance0601} by adding a simple fermion field term to the Lagrangian density.  The term represents the kinetic energy density of an O1 electron relative to the expectation value of the position of O1's centre of mass. It is postulated that quantum indeterminacy involved in transformations between O2 and O1's reference systems (and possibly within each reference system) may effectively require, or be equivalent to a standard theory with, $d$ spinor fields $\psi_{\mu}$ instead of the single field $\psi$, where $d$ is the number of spacetime dimensions. The term $i\bar\psi^{\mu} \gamma ^{\alpha}\partial_{\alpha} \psi_{\mu}$ is then added to O1's centre of mass kinetic energy (density) to give the Lagrangian density.  This Lagrangian density looks similar to that of a superstring theory, if the O1 coordinates ${x^\alpha}$ are restricted.  As in~\cite{Dance0601}, this restriction is achieved by postulating that O1 is unable to relay information about the $r$ coordinate, and by postulating that O1 cannot measure coordinates, for its own state, with respect to which its quantum state remains symmetric.  The end result is a Lagrangian density resembling that of superstring theory.  It is tentatively suggested that a basis for superstring/M theory may be the dynamics and information processing capacities of such an observer O1, if the postulates in this line of argument hold.


\begin{thebibliography}{9}
 
\bibitem{Schwarz0702} J. H. Schwarz, "String Theory: Progress and Problems", hep-th/0702219. 
 
\bibitem{Dance0601} "Symmetry limitations on quantum mechanical observers, and conjectured link with string theory", hep-th/0601104.
 
\bibitem{Bartlett0610}  S. D. Bartlett,  T. Rudolph and R. W. Spekkens, "Reference frames, superselection rules, and quantum information", Rev. Mod. Phys. 79, 555 (2007).
 
\bibitem{Son0804} D.T. Son, "Toward an AdS/cold atoms correspondence: a geometric realization of the Schroedinger symmetry", Phys. Rev. D 78, 046003 (2008).
 
\bibitem{BM0804}  K. Balasubramanian and J. McGreevy, "Gravity duals for non-relativistic CFTs", Phys. Rev. Lett.101:061601 (2008)
 
\bibitem{Merkl0803} M. Merkl, F.E. Zimmer, G. Juzeliunas, and P. Ohberg, "Atomic Zitterbewegung", arXiv:0803.4189.
 
\bibitem{Wen03} Xiao-Gang Wen, "Quantum order from string-net condensations and origin of light and massless fermions", Phys. Rev. D68, 024501 (2003) 
 
\bibitem{Gu&Wen06} Z. Gu and X. Wen, "A lattice bosonic model as a quantum theory of gravity", gr-qc/0606100.
 
\bibitem{Bohr2004a} A. Bohr, B.R. Mottelson, and O. Ulfbeck, "The Principle Underlying Quantum Mechanics", Found. Phys. 34, p405 (2004). 
 
\bibitem{Bohr2004b} A. Bohr, B.R. Mottelson, and O. Ulfbeck, "Quantum World is Only Smoke and Mirrors", Physics Today 57/10 (2004). 
 
\bibitem{Wang0609} W. Wang, "Entanglement and Disentanglement, Probabilistic Interpretation of Statevectors, and Transformation between Intrinsic Frames of Reference", quant-ph/0609093.
 
\bibitem{Dance0610} M. Dance, "Quantum mechanical transformation between reference frames - a discursive spacetime diagram approach", physics/0610234.
 
\bibitem{heisenberg}W. Heisenberg, Zeitschrift fur Physik, 43, 172-98 (1927). English translation reproduced in "Quantum Theory and Measurement", ed. J. Wheeler and W.H. Zurek,  Princeton University Press, 
1983.
 
\bibitem{Kim&Noz0609} Y.S. Kim and M.E. Noz, "Can you do quantum mechanics without Einstein?", quant-ph/0609127.
 
\bibitem{hepth0805234}  S. Banerjee and A. Sen, " Interpreting the M2-brane Action", arXiv:0805.3930. 
 
\bibitem{hep-ex/0008017} J. Schwarz, "An introduction to superstring theory for experimentalists", hep-ex/0008017. 
 
\bibitem{Kim&Noz08} Y.S.Kim and M.E.Noz, "Can the quark model be relativistic enough to include the parton model?", chapter in edited volume "New Ideas in the Quark Model" (Nova Publishing), arXiv:0803.2633.
  
\end{thebibliography}
\end{document}